\documentclass[prd,amsfonts,onecolumn,superscriptaddress,aps,nofootinbib,12pt]{revtex4-1}

\usepackage{amsmath, amsthm, amssymb, amsfonts, mathrsfs}
\usepackage{physics, braket, slashed, siunitx, bm}
\usepackage{bbold}
\usepackage{listings}
\usepackage{color}
\usepackage[usenames,dvipsnames]{xcolor}
\usepackage{graphicx}
\usepackage{float}
\usepackage{placeins}
\usepackage{booktabs}
\usepackage{hyperref}
\usepackage{soul}
\usepackage{ulem}
\def\ds{\partial\!\!\!/}
\def\As{A\!\!\!/}
\def\bs{b\!\!\!/}
\def\ks{k\!\!\!/}
\def\ps{p\!\!\!/}
\def\us{u\!\!\!/}

\newcommand{\bea}{\begin{eqnarray}}
	\newcommand{\eea}{\end{eqnarray}}

\begin{document}

\title{On aether terms in a space-time with a compact extra dimension}
\author{J. Furtado}
\email{job.furtado@ufca.edu.br}
\affiliation{Centro de Ci\^{e}ncias e Tecnologia, Universidade Federal do Cariri\\
63048-080, Juazeiro do Norte, Cear\'{a}, Brazil}
\author{T. Mariz}
\email{tmariz@fis.ufal.br} 
\affiliation{Instituto de F\'\i sica, Universidade Federal de Alagoas, 57072-270, Macei\'o, Alagoas, Brazil}
\author{J. R. Nascimento}
\email{jroberto@fisica.ufpb.br}
\affiliation{Departamento de F\'{\i}sica, Universidade Federal da Para\'{\i}ba\\
 Caixa Postal 5008, 58051-970, Jo\~ao Pessoa, Para\'{\i}ba, Brazil}
\author{A. Yu. Petrov}
\email{petrov@fisica.ufpb.br}
\affiliation{Departamento de F\'{\i}sica, Universidade Federal da Para\'{\i}ba\\
 Caixa Postal 5008, 58051-970, Jo\~ao Pessoa, Para\'{\i}ba, Brazil}
 
\begin{abstract}
In this paper, we explicitly calculate the aether-like corrections for the electromagnetic field in the case when the space-time involves an extra compact spatial dimension besides of usual four dimensions. Our methodology is based on an explicit summation over the Kaluza-Klein tower of fields which turns out to be no more difficult than the finite-temperature calculations. We found some new terms which can be treated as certain five-dimensional generalizations of the aether-like terms $\kappa^{mnpq}F_{mn}F_{pq}$ and $\bar{\psi}c^{mn}\gamma_m\partial_n\psi$ which in four dimensions are treated as important ingredients of the Lorentz-breaking extension of the standard model. We have shown that these new terms become to dominate in the limit where the extra dimension is very small, allowing us to suggest that there will be essential modifications of quantum corrections if our space indeed has a small extra spatial dimension. Moreover, we have shown that in the CPT-even case we have extra new particles as a consequence of the extra dimension.
\end{abstract}

\maketitle

\section{Introduction}
The idea of compact extra dimensions has a long history. Originally, it has been proposed as a Kaluza-Klein theory \cite{KK,KK1} aimed to unify gravity and electromagnetism. Being abandoned for a long time, this idea has been revitalized with development of string theory known to be well defined in space-times with dimensions much larger than four (for a review on applications of the Kaluza-Klein concept in modern contexts, including cosmology, phenomenology of elementary particles, and ten- and eleven-dimensional supergravity, see e.g. \cite{BL,ArkHam}). An important step in development of this concept has been made in a seminal paper \cite{Carr} where it has been immersed into the Lorentz-breaking context (another approach to this problem has been discussed in \cite{Rizzo}). Following \cite{Carr}, if the Lorentz-breaking vectors are directed along the compact extra dimensions, the masses of the particles will be modified, probably affecting the hierarchy problem (some important results in studies of an impact of extra compact dimension or quantum corrections were obtained in \cite{Yamashita}, where the two-point function of the massless QED, although without explicit Lorentz-breaking terms in the action, has been calculated in a five-dimensional space, with a compact extra dimension). Therefore, the natural problem is -- how the modifications of the known Lorentz-breaking results will behave if one (extra) dimension is compactified, and how the presence of the compact extra dimension will affect the known results? The main motivation for this study is the following one: it is known that the compactness of one of dimensions in some field theory sometimes can yield large modifications of quantum corrections. For example, it is shown in \cite{CST} that, if we consider the gravitational Chern-Simons term at the finite temperature -- we note that the calculation scheme for the finite temperature is just the same as in the case of the compact extra dimension, so that the high temperature situation is a perfect analogue of the presence of the small extra dimension -- we see that the result for this term grows as square of temperature. Therefore studying quantum corrections for the case of the compact extra dimension could allow us, in principle, to make estimations of the size of this dimension. 
In principle, we can consider a situation more generic than that one discussed in \cite{Carr}, and suggest that the Lorentz-breaking vectors are arbitrary, but one of spatial dimensions is compact. 

Up to now, there are a few examples of studies of Lorentz-breaking theories in a five-dimensional space-time (besides of the papers \cite{Carr,Rizzo}, one can also mention \cite{Obousy}). One of these examples is presented in the paper \cite{aether} by some of us, where the aether terms (both for gauge and spinor fields) have been generated from a nonminimal (magnetic), CPT-odd coupling of the gauge field. Moreover, compactness of the extra dimension certainly implies modification of the result for the aether terms in comparison with \cite{aether}, so, it is interesting to generalize the result of \cite{aether} within this scenario. In this paper, we perform this calculation. Besides of this, present another scheme to generate the aether-like term in five dimensions, that is, extend the result of \cite{Maluf}, where the Lorentz symmetry breaking is introduced through the additive CPT-even term, for the case when one extra compact dimension is added. The methodology of calculations is apparently very similar to the finite-temperature field theory, where we also assume the projections of momenta along a certain dimension to be discrete, thus, we can use all well-developed machinery of the finite-temperature field theory.

The structure of the paper looks like follows. In the section 2, we calculate the lower CPT-even Lorentz-breaking contribution to the two-point function of the gauge field for two possible Lorentz-breaking extensions of the five-dimensional QED. In the section 3 we calculate the two-point function of the spinor field in these theories. The section 4 is the Summary where our results are discussed.

\section{Two-point function of the gauge field}

To start the calculation, we write down the extended Lorentz-breaking QED in five dimensions involving the new coupling \cite{aether}:
\bea
\label{odd}
S_1=\int d^5x\Big(\bar{\psi}(i\ds-m-e\As-g\epsilon^{mnpqr}b_mF_{np}\sigma_{qr})\psi-\frac{1}{4}F_{mn}F^{mn}\big).
\eea
For definiteness, we consider $b_m=(b_\mu,b_5)$, where $\mu=0,1,2,3$, and so on. We note that here there is no need to introduce the aether-like term at the very beginning since the quantum aether-like contribution is explicitly finite in five dimensions due to the magic of dimensional regularization.

As a first step, we calculate the two-point function of the gauge field. The contribution generated by two minimal vertices has been discussed in great detail in \cite{Yamashita}. It does not depend on Lorentz-breaking parameters. The mixed contribution, involving both minimal and nonminimal vertices, is
\bea
\Pi_{mixed}=4meg\epsilon^{abcde}b_aF_{bc}F_{de}\int\frac{d^5k}{(2\pi)^5}\frac{1}{(k^2-m^2)^2}.
\eea
However, this expression evidently represents itself as a total derivative (the $\epsilon^{abcde}b_aF_{bc}F_{de}$ is a perfect analogue of the $4D$ topological term $\epsilon^{bcde}F_{bc}F_{de}$, and, moreover, if we choose $b_5$ to be along the extra dimension as we do below, we will explicitly reproduce the four-dimensional topological term $\tilde{F}F$) independently on whether the extra dimension is compact or not (this is a difference from the four-dimensional case where the mixed contribution yields the CFJ term \cite{Scarp}), so this term can be disregarded. 

So, we begin our study with the contribution generated by two non-minimal vertices, that is, the aether-like correction, depicted in Fig.~\ref{F1}, which is given by
\bea
\Pi_1=\frac{g^2}{2}\epsilon^{mnpqr}b_mF_{np}\epsilon^{m'n'p'q'r'}b_{m'}F_{n'p'}{\rm tr}\int\frac{d^5k}{(2\pi)^5}\sigma_{qr}S(k)\sigma_{q'r'}S(k),
\eea
where
\bea
S(k)=i(\ks-m)^{-1},
\eea
with $\slashed{k}=k_a\gamma^a$, is the usual propagator of the spinor field.

\vspace*{2mm}

\begin{figure}[ht]
\centerline{\includegraphics{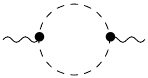}} 
\caption{First-order Lorentz-breaking contributions.}
\end{figure}
\label{F1}

\vspace*{2mm}

Taking into account only the relevant (that is, even with respect to $k$) terms, we find that this correction is
\bea
\Pi_1=\Pi_1^{(1)}+\Pi_2^{(2)},
\eea
where
\bea
\Pi_1^{(1)}&=&-\frac{m^2g^2}{2}\epsilon^{mnpqr}b_mF_{np}\epsilon^{m'n'p'q'r'}b_{m'}F_{n'p'}{\rm tr}(\sigma_{qr}\sigma_{q'r'})\int\frac{d^5k}{(2\pi)^5}\frac{1}{(k^2-m^2)^2},\nonumber\\
\Pi_1^{(2)}&=&-\frac{g^2}{2}\epsilon^{mnpqr}b_mF_{np}\epsilon^{m'n'p'q'r'}b_{m'}F_{n'p'}{\rm tr}(\gamma_l\sigma_{qr}\gamma_{l'}\sigma_{q'r'})\int\frac{d^5k}{(2\pi)^5}\frac{k^lk^{l'}}{(k^2-m^2)^2}.
\eea
We can rewrite these expressions as
\bea
\Pi_1^{(1)}&=&-\frac{m^2g^2}{2}\epsilon^{mnpqr}b_mF_{np}\epsilon^{m'n'p'q'r'}b_{m'}F_{n'p'}{\rm tr}(\sigma_{qr}\sigma_{q'r'})I_1, \\
\Pi_1^{(2)}&=&-\frac{g^2}{2}\epsilon^{mnpqr}b_mF_{np}\epsilon^{m'n'p'q'r'}b_{m'}F_{n'p'}{\rm tr}(\gamma_l\sigma_{qr}\gamma_{l'}\sigma_{q'r'})I_{1'}^{ll'},
\eea
with
\bea
I_1&=&\int\frac{d^5k}{(2\pi)^5}\frac{1}{(k^2-m^2)^2}, \\
I_{1}^{ll'}&=&\int\frac{d^5k}{(2\pi)^5}\frac{k^lk^{l'}}{(k^2-m^2)^2}.
\eea

Now, let us suggest that our fifth extra dimension is compact, of the size $L$. It means that actually we can proceed with Matsubara formalism, where the momentum component corresponding to the extra dimension is $k^5=\frac{2\pi n}{L}=-k_5$, with $n$ being an integer, to be compatible with the periodicity of this dimension (and not semi-integer).  Afterwards, carrying out the Wick rotation with respect to $k_0$, i.e., by considering $k_0=ik_{0E}$, and introducing $\bar{k}^2=k^2_{0E}+k^2_1+k^2_2+k^2_3$, we write
\bea \label{I12}
I_1&=&\frac{i}{L}\sum\limits_{n=-\infty}^{\infty}\int\frac{d^4\bar{k}}{(2\pi)^4}\frac{1}{(\bar{k}^2+k_5^2+m^2)^2}, \nonumber\\
I_{1'}^{ll'}&=&\frac{i}{L}\sum\limits_{n=-\infty}^{\infty}\int\frac{d^4\bar{k}}{(2\pi)^4}\frac{\frac{\bar k^2}{4}(\delta^{ll'}-t^lt^{l'})+k_5^2t^lt^{l'}}{(\bar{k}^2+k_5^2+m^2)^2},
\eea
where we have considered the decomposition $k^l=\bar k^l+k^5t^l$,  with $\bar k^l=(k^\lambda,0)$ and $t^l=(0,0,0,0,1)$, and the substitution $\bar k^l \bar k^{l'} \to \frac{\bar k^2}{4} (\delta^{ll'}-t^l t^{l'})$.  

In order to calculate the above sum-integrals, initially, in (\ref{I12}), we evaluate the integrals in $d$ dimensions, so that $d^4\bar{k}/(2\pi)^4$ goes to $\mu^{4-d}d^d\bar{k}/(2\pi)^d$. The result is
\bea
I_1&=&\frac{i}{L}\frac{\mu^{4-d}}{(4\pi)^{d/2}}\Gamma\left(2-\frac{d}{2}\right)\sum\limits_{n=-\infty}^{\infty}\frac{1}{(k_5^2+m^2)^{2-d/2}},\nonumber\\
I_{1'}^{\alpha\beta}&=&\frac{i}{2L}\frac{\mu^{4-d}}{(4\pi)^{d/2}}\Gamma\left(1-\frac{d}{2}\right)\sum\limits_{n=-\infty}^{\infty}\frac{\delta^{\alpha\beta}}{(k_5^2+m^2)^{1-d/2}}, \nonumber\\
I_{1'}^{55}&=&\frac{i}{L}\frac{\mu^{4-d}}{(4\pi)^{d/2}}\Gamma\left(2-\frac{d}{2}\right)\sum\limits_{n=-\infty}^{\infty}\frac{k_5^2}{(k_5^2+m^2)^{2-d/2}}.
\eea
Finally, to calculate the the sum over the Matsubara frequencies, we use the formula \cite{Ford}
\begin{equation}\label{sum}
\sum\limits_{n=-\infty}^{\infty}[(n+\eta)^2+\xi^2]^{-\lambda} = \frac{\sqrt{\pi}\Gamma(\lambda-1/2)}{\Gamma(\lambda)(\xi^2)^{\lambda-1/2}}+4\sin(\pi\lambda)f_\lambda(\eta,\xi),
\end{equation}
with
\begin{equation}\label{f}
f_\lambda(\eta,\xi) = \int^{\infty}_{|\xi|}\frac{dz}{(z^2-\xi^2)^{\lambda}}Re\left(\frac{1}{e^{2\pi(z+i\eta)}-1}\right).
\end{equation}
The above expression is valid only for $\lambda<1$, aside from the poles at $\lambda=1/2,-1/2,-3/2,\cdots$. However, this restriction can be circumvented when we use the recurrence relation
\bea
f_{\lambda}(\eta,\xi) &=& -\frac1{2\xi^2}\frac{2\lambda-3}{\lambda-1}f_{\lambda-1}(\eta,\xi) - \frac1{4\xi^2}\frac1{(\lambda-2)(\lambda-1)}\frac{\partial^2}{\partial\eta^2}f_{\lambda-2}(\eta,\xi).
\eea
Note that, in our case, one has $\lambda=1-d/2$ and $\lambda=2-d/2$, so that (\ref{sum}) can be used directly. Taking all together, with $\eta=0$ and $\xi=\frac{mL}{2\pi}$, we arrive at
\bea\label{I12a}
I_1 &=& - \frac{im}{16\pi^2} + \frac{i}{8\pi L} \int_{|\xi|}^{\infty}dz\ (\coth(\pi z)-1), \nonumber\\
I_{1'}^{\alpha\beta}&=& \frac{im^3}{48\pi^2}\delta^{\alpha\beta} + \frac{i\pi}{4L^3}\delta^{\alpha\beta} \int_{|\xi|}^{\infty}dz\ (z^2-\xi^2)(\coth(\pi z)-1), \nonumber\\
I_{1'}^{55}&=& \frac{im^3}{24\pi^2} - \frac{i\pi}{4L^3} \int_{|\xi|}^{\infty}dz\ (z^2+\xi^2)(\coth(\pi z)-1),
\eea
where we have taken into account an expansion around $d=4$. We note however that the finiteness of these integrals is a consequence of use of the dimensional regularization (which gives the finite result for Euler gamma function of negative fractional arguments in (\ref{sum})). Had we use other regularization, these contributions could in principle display pole parts as well, however, the dimensional regularization is the most appropriate one for calculations within the compactification process. 

Thus, by using the first solution of~(\ref{I12a}) in $\Pi_1^{(1)}$, after the inverse Wick rotation, we obtain
\bea
\Pi_1^{(1)}&=& \left[\frac{m^3g^2}{\pi^2}-\frac{8\pi g^2}{L^3}\int_{|\xi|}^{\infty}dz\ \xi^2 (\coth(\pi z)-1)\right](b^2F_{mn}F^{mn}-2b_nF^{mn}b^lF_{ml}),
\eea
which matches the zero temperature result found in \cite{aether} (where the Lorentz invariant part proportional to $b^2$ was omitted) for this sector of the contribution.

The contribution $\Pi_1^{(2)}$ would have a much more complicated form since $I_2^{ll'}$ is no more proportional to $\delta^{ll'}$. Explicitly, it looks like
\bea
\Pi_1^{(2)}=-\frac{g^2}{2}\epsilon^{mnpqr}b_mF_{np}\epsilon^{m'n'p'q'r'}b_{m'}F_{n'p'}\left[{\rm tr}(\gamma_\alpha\sigma_{qr}\gamma_\beta\sigma_{q'r'})I_2^{\alpha\beta} + {\rm tr}(\gamma_4\sigma_{qr}\gamma_4\sigma_{q'r'})I_2^{55}\right].
\eea
Since $I_2^{\alpha\beta}$ and $I_2^{55}$ are not mutually proportional, we see that in the case of finite $L$, their explicit $L$-dependent contributions will be essentially different. Then, the result takes the form:
\bea
\Pi_1^{(2)}&=&-\left[\frac{m^3g^2}{3\pi^2}-\frac{8\pi g^2}{L^3}\int_{|\xi|}^{\infty}dz\ (\xi^2-2z^2) (\coth(\pi z)-1)\right](b^2F_{mn}F^{mn}-2b_nF^{mn}b^lF_{ml})  \nonumber\\
&&+\left[\frac{8\pi g^2}{L^3}\int_{|\xi|}^{\infty}dz\ (\xi^2-3z^2) (\coth(\pi z)-1)\right](b_5^2F_{mn}F^{mn}+2b_mF^{m5}b^nF_{n5} \nonumber\\
&&-2b^2F_{m5}F^{m5}-4b_nF^{mn}b_5F_{m5}).
\eea
The final result is a sum of these two contributions, $\Pi_1=\Pi_1^{(1)}+\Pi_1^{(2)}$, given by
\bea
\Pi_1&=&\left[\frac{2m^3}{3\pi^2}-\frac{16\pi F_1(\xi)}{L^3}\right]g^2(b^2F_{mn}F^{mn}-2b_nF^{mn}b^lF_{ml})\\
&&+\frac{8\pi G_1(\xi)}{L^3}g^2(b_5^2F_{mn}F^{mn}+2b_mF^{m5}b^nF_{n5}-2b^2F_{m5}F^{m5}-4b_nF^{mn}b_5F_{m5}), \nonumber
\eea
where
\bea\label{f12}
F_1(\xi)&=&\int_{|\xi|}^{\infty}dz\ z^2 (\coth(\pi z)-1),\nonumber\\
G_1(\xi)&=&\int_{|\xi|}^{\infty}dz\ (\xi^2-3z^2) (\coth(\pi z)-1).
\eea
It is easy to verify that in $L\to\infty$ (or $\xi\to\infty$) limit, this result matches the expression obtained in \cite{aether} (where the terms proportional to $b^2$ were not discussed). At the same time, for the small $L$, the $L\to 0$ impact begins to dominate, so that the integrals (\ref{f12}) assume their asymptotic form, given by 
\bea
F_1(\xi\to0)&=&\int_{0}^{\infty}dz\ z^2 (\coth(\pi z)-1) = \frac{\zeta(3)}{2\pi^3},\nonumber\\
G_1(\xi\to0)&=&-\int_{0}^{\infty}dz\ 3 z^2 (\coth(\pi z)-1) = -\frac{3\zeta(3)}{2\pi^3}.
\eea

It is interesting to consider the following example. Let us suggest that the Lorentz-breaking vector $b_m$ is directed along the extra dimension, looking like $b_m=(0,0,0,0,b_5)$, so, $b^2=-b_5^2$. In this case, we have
\bea\label{Pi1}
\Pi_1&=&-\left[\frac{2m^3}{3\pi^2}-\frac{16\pi F_1(\xi)}{L^3}-\frac{8\pi G_1(\xi)}{L^3}\right]g^2b_5^2F_{\mu\nu}F^{\mu\nu}.
\eea
We note that in this case the fifth dimension manifests itself only through the modification of quantum corrections for the usual four-dimensional photons, with there is no $F_{\alpha5}$ components in this contribution and hence no new particles corresponding to the extra dimension. We note that this situation is phenomenologically more advantageous since it explains the absence of new particles originated from extra dimensions, and, at the same time, allows to have significant modifications of the Maxwell term. Actually, this term suffers a large finite renormalization, with a coefficient dependent on the size of the extra dimension. The contributions involving $F_1(\xi)$ and $G_1(\xi)$ are the corrections which are absent in the case of the infinite-size extra dimension. Actually, we found that its finiteness would correct the aether term, as more, as the size of the extra dimension is less. Moreover, if the size of the extra dimension, the $L$, is enough small,  we find that the new contributions will dominate, hence the small extra dimension will yield a large quantum correction which can be of the compatible or even higher order with the usual Maxwell-like correction, dependently on the relation between $gb_5$ and $L$.

\begin{figure}[ht]
\centerline{\includegraphics{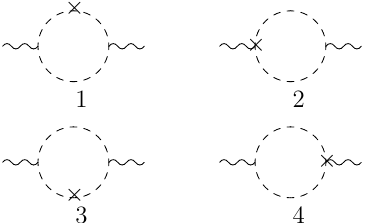}} 
 \caption{First-order Lorentz-breaking contributions.}
 \label{F2}
\end{figure}

Another possibility for including the Lorentz symmetry breaking in the vector-spinor coupling would consist in introducing the CPT-even Lorentz-breaking term into the action as it has been done in \cite{Maluf}. For this, we must consider the action given by
\bea
\label{even}
S_2 =\int d^5x\Big[\bar{\psi}(i\slashed{\partial}-m+i c^{ab} \gamma_a \partial_b-e\slashed{A}-e c^{ab} \gamma_a A_b)\psi-\frac{1}{4}F_{ab}F^{ab}-\frac{1}{4}\kappa_{abcd}F^{ab}F^{cd}\Big].
\eea
Here the $\kappa_{abcd}$ is the constant fourth-rank tensor possessing the same symmetry as the Riemann curvature tensor. In the simplest case, which we use within this paper, we use its following definition: $\kappa_{abcd}=\eta_{ac}u_bu_d-\eta_{ad}u_bu_c+\eta_{bd}u_au_c-\eta_{bc}u_au_d$. In this case the LV term $-\frac{1}{4}\kappa_{abcd}F^{ab}F^{cd}$ is reduced to the aether-like form $u^au_bF_{ac}F^{bc}$.
Thus, we have
\bea
\Pi_2 = \frac12 A_m \Pi_2^{mn} A_n,
\eea
where $\Pi_2^{mn}=\Pi_{2,1}^{mn}+\Pi_{2,2}^{mn}+\Pi_{2,3}^{mn}+\Pi_{2,4}^{mn}$, with
\begin{eqnarray}
\Pi_{2,1}^{mn} &=& -ie^2\mathrm{tr} \int\frac{d^{5}p}{(2\pi)^5}S(p)u^au^b\gamma_ap_bS(p)\gamma^mS(p-i\partial)\gamma^n,\\
\Pi_{2,2}^{mn} &=& ie^2\mathrm{tr} \int\frac{d^{5}p}{(2\pi)^5}S(p)u^mu^a\gamma_a S(p-i\partial)\gamma^n,\\
\Pi_{2,3}^{mn} &=& -ie^2\mathrm{tr} \int\frac{d^{5}p}{(2\pi)^5}S(p)\gamma^m S(p-i\partial)u^au^b\gamma_a(p-i\partial)_bS(p-i\partial) \gamma^n,\\
\Pi_{2,4}^{mn} &=& ie^2\mathrm{tr} \int\frac{d^{5}p}{(2\pi)^5}S(p)\gamma^mS(p-i\partial)u^nu^a\gamma_a.
\end{eqnarray}
For simplicity, we have considered $c_{ab}=u_a u_b$, with $u_a=(u_\alpha,u_5)$. These contributions are depicted in Fig.~\ref{F2}.

Supposing that the extra dimension again is compact, proceeding with the Matsubara formalism, we can present our result as a sum $\Pi_2=\Pi_2^{(1)}+\Pi_2^{(2)}+\Pi_2^{(3)}$, where
\bea
\Pi_2^{(1)} &=& \left[\frac{m}{24\pi^2}-\frac{F_2(\xi)}{12\pi L}\right]e^2A_m\left[\left(2 (u\cdot \partial)^2-u^2 \Box\right) g^{mn}+u^m \left(2 \partial^2 u^n-2 \partial^n (u\cdot \partial)\right)\right. \nonumber\\
&&\left.+\partial^m \left(u^2 \partial^n-2 u^n (u\cdot \partial)\right)\right]A_n,\nonumber\\
\Pi_2^{(2)} &=& \frac{\pi G_2(\xi)}{24L}e^2A_m\left[t^m \left(t^n \left(2 (u\cdot\partial)^2-u^2 \partial^2\right)+(\partial\cdot t) \left(u^2 \partial^n-2 u^n u\cdot\partial\right)\right)\right.\nonumber\\ 
&&+\left. t^n (\partial\cdot t) \left(u^2 \partial^m-2 u^m u\cdot\partial\right)+(\partial\cdot t)^2 \left(2 u^m u^n-u^2 g^{mn}\right) \right.\nonumber\\
&&+\left. 2 (u\cdot t)^2 \left(\partial^2 g^{mn}-\partial^m \partial^n\right)\right] A_n,\nonumber\\
\Pi_2^{(3)} &=& \frac{\pi H_2(\xi)}{24L}e^2A_m(u\cdot t)^2 \left(g^{mn} (\partial\cdot t)^2+\partial^2 t^m t^n-(\partial\cdot t) \left(\partial^n t^m+\partial^m t^n\right)\right) A_n,
\eea
with
\bea\label{FGH}
F_2(\xi)&=&\int_{|\xi|}^{\infty}dz\ (\coth(\pi z)-1), \nonumber\\
G_2(\xi)&=&\int_{|\xi|}^{\infty}dz\ (\xi^2-z^2) \coth(\pi z)\mathrm{csch}^2(\pi z), \nonumber\\
H_2(\xi)&=&\int_{|\xi|}^{\infty}dz\ (\xi^2-3z^2) \coth(\pi z)\mathrm{csch}^2(\pi z).
\eea
Again, the $t^m$ is a fictitious unit vector along the extra dimension. It is easy to see that all these three contributions are gauge invariant and can be presented as
\bea
\Pi_2^{(1)} &=& (4u^mF_{mn}u_lF^{ln}-u^2F_{mn}F^{mn}) \left(\frac{e^2m}{48\pi^2}-\frac{e^2}{24\pi L}F_2(\xi)\right),\nonumber\\
\Pi_2^{(2)} &=& \frac12 \left[-F_{mn}F^{mn}(u\cdot t)^2+u^2t^mF_{mn}t_lF^{ln}-2(t^mu^lF_{ml})^2
\right]\nonumber\\
&&\times\frac{e^2\pi}{12L}G_2(\xi),\nonumber\\
\Pi_2^{(3)} &=& \frac12  (u\cdot t)^2t^mF_{mn}t_lF^{ln}  
\frac{e^2\pi}{12L}H_2(\xi).
\eea
We note that only the first of them does not vanish if the extra dimension becomes large. 

We note that if we suggest the Lorentz-breaking vector $u^m$ to be directed along the extra dimension as well, so, $u_m=\lambda t_m$, with $\lambda$ is a some number, our results would take the form
\bea
\Pi_2^{(1)} &=& \lambda^2(\frac{3}{2}F_{5\alpha}F^{5\alpha}+\frac{1}{4}F_{\alpha\beta}F^{\alpha\beta}) 
\left[\frac{e^2m}{12\pi^2}-\frac{e^2}{6\pi L}F_2(\xi)\right],\nonumber\\
\Pi_2^{(2)} &=& -\frac{\lambda^2}{2} \left[F_{\alpha\beta}F^{\alpha\beta}+3F_{5\alpha}F^{5\alpha}\right]
\frac{e^2\pi}{12L}G_2(\xi),\nonumber\\
\Pi_2^{(3)} &=& \frac{\lambda^2}{2} F_{5\alpha}F^{5\alpha}  
\frac{e^2\pi}{12L}H_2(\xi).
\eea
Let us again, as above, assume $b_{\mu}$ to be directed along the extra dimension. In this case we can write the sum of these expressions as
\bea
\Pi_2&=&\lambda^2F_{\alpha\beta}F^{\alpha\beta}\Big[
\frac{1}{4}\left[\frac{e^2m}{12\pi^2}-\frac{e^2}{6\pi L}F_2(\xi)\right]-\frac{1}{2}\frac{e^2\pi}{12L}G_2(\xi)
\Big]+\nonumber\\&+&
\lambda^2F_{5\alpha}F^{5\alpha}\Big[
\frac{3}{2}\left[\frac{e^2m}{12\pi^2}-\frac{e^2}{6\pi L}F_2(\xi)\right]-
\frac{3}{2}\frac{e^2\pi}{12L}G_2(\xi)+\frac{1}{2}\frac{e^2\pi}{12L}H_2(\xi)
\Big].
\eea
We see that in this case we have contributions both of usual four-dimensional photons and of new particles arising due to the presence of the extra dimension. For the small $L$, the terms with $\frac{1}{L}$ become dominant, so the integrals (\ref{FGH}) assume their asymptotic form:
\bea
F_2(\xi\to0)&=&\int_{0}^{\infty}dz\ (\coth(\pi z)-1) \approx 11.1, \nonumber\\
G_2(\xi\to0)&=&-\int_{0}^{\infty}dz\ z^2 \coth(\pi z)\mathrm{csch}^2(\pi z) \approx -1.2, \nonumber\\
H_2(\xi\to0)&=&-\int_{0}^{\infty}dz\ 3z^2 \coth(\pi z)\mathrm{csch}^2(\pi z) \approx -3.6.
\eea
Again, we find that the new contributions, depending on the size of the extra dimension $L$, are proportional to $L^{-1}$ as in the previous case.

We close this section by the following conclusion: we found that, for both schemes of introducing the Lorentz symmetry breaking, CPT-odd and CPT-even ones, the two-point function is composed by a sum of the usual result corresponding to the infinite extra dimension, and additive terms which strongly increase if the size of the extra dimension is small. Effectively it means that the presence of the microscopic extra dimension modifies strongly the kinetic term for the gauge field increasing as in inverse size of the extra dimension. Besides, we note that in certain cases, namely, for the CPT-odd scheme for the Lorentz symmetry breaking with the LV vector aligned with the extra dimension, there is no contribution described by the extra dimension particles.

\section{Two-point function of the spinor field}

While the gauge sector attracted the most attention, it is nevertheless interesting to study as well the contributions to the two-point spinor function generalizing the studies carried out in \cite{KostMew} to the five-dimensional case. To proceed with contributions including external spinor fields, it is important to list some standard five-dimensional sum-integrals. One of them is the $I_1$ calculated above (\ref{I12a}). The other integrals will emerge in the sequel. Using the Matsubara formalism, we find
\bea
\label{list}
I_1(m^2) &=& \int\frac{d^5k}{(2\pi)^5}\frac{1}{(k^2-m^2)^2}\to\frac{i}{L}\sum\limits_{n=-\infty}^{\infty}\int\frac{d^4\bar{k}}{(2\pi)^4}\frac{1}{(\bar{k}^2+4\frac{\pi^2}{L^2}n^2+m^2)^2}=\nonumber\\
&=&- \frac{im}{16\pi^2} + \frac{i}{16\pi L}F_2(\xi); \nonumber\\
I_2(m^2) &=& \int\frac{d^5k}{(2\pi)^5}\frac{1}{k^2(k^2-m^2)^2}\to\frac{2i}{L}\sum\limits_{n=-\infty}^{\infty}\int_0^1dx\int\frac{d^4\bar{k}}{(2\pi)^4}\frac{1}{(\bar{k}^2+4\frac{\pi^2}{L^2}n^2+m^2x)^3}=\nonumber\\
&=& 2\int_0^1dx I_3(m^2x)=-\frac{i}{8\pi^2m};\nonumber\\
I_3(m^2) &=& \int\frac{d^5k}{(2\pi)^5}\frac{1}{(k^2-m^2)^3}\to-\frac{i}{L}\sum\limits_{n=-\infty}^{\infty}\int\frac{d^4\bar{k}}{(2\pi)^4}\frac{1}{(\bar{k}^2+4\frac{\pi^2}{L^2}n^2+m^2)^3}=-\frac{i}{32\pi^2m};\nonumber\\
I_4(m^2) &=& \int\frac{d^5k}{(2\pi)^5}\frac{1}{k^2(k^2-m^2)}\to\frac{i}{L}\sum\limits_{n=-\infty}^{\infty}\int_0^1dx\int\frac{d^4\bar{k}}{(2\pi)^4}\frac{1}{(\bar{k}^2+4\frac{\pi^2}{L^2}n^2+m^2x)^2}=\nonumber\\
&=&-\frac{im}{48\pi^2}+\frac{i}{8\pi L}\int_0^1 dx F_2(\xi\sqrt{x})
;\nonumber\\
I_5(m^2) &=& \int\frac{d^5k}{(2\pi)^5}\frac{k^2}{(k^2-m^2)^3}=I_1(m^2)+m^2I_3(m^2)=\nonumber\\
&=&- \frac{3im}{32\pi^2} + \frac{i}{8\pi L} F_2(\xi);\nonumber\\
I_6(m^2)  &=&\int\frac{d^5k}{(2\pi)^5}\frac{k^2}{(k^2-m^2)^2}=\int\frac{d^5k}{(2\pi)^5}\frac{1}{k^2-m^2}-m^2I_1(m^2)=\nonumber\\
&=&\frac{im^3}{24\pi^2}-\frac{i}{4\pi L^3}(F_1(\xi)-\xi^2F_2 (\xi)).
\eea
It is interesting to note that $I_2$, and hence $I_3$, is $L$-independent.

\begin{figure}[htbp]
\centerline{\includegraphics{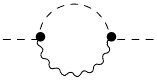}} 
\caption{Contribution to the two-point function of the spinor field.}
\label{F3}
\end{figure}

In this case, it is interesting to consider the two-point function of the spinor field. It is generated by the Feynman diagram given by Fig.~\ref{F3}, its purely nonminimal contribution, that is, the lower CPT-even one, in the Feynman gauge, is given by the following expression:
\bea
\Sigma_1=4g^2\int\frac{d^5k}{(2\pi)^5}\bar{\psi}(-p)\epsilon^{abcde}\sigma_{de}
\frac{[\gamma^{f}(k_{f}+p_{f})-m]}{[(k+p)^2-m^2]}\epsilon^{a'b'c'd'e'}\sigma_{d'e'}
\eta_{cc'}b_{a}b_{a'}
\frac{k_{b}k_{b'}}{k^2}\psi(p).
\eea

\vspace*{2mm}

\vspace*{2mm}

We note that, in order to have the CPT-even, aether-like term for the spinor field, we should have just two insertions of the first order in $b_a$, since the desired term is proportional to $b_ab_b$ \cite{Carr}. So, let us proceed with this contribution.  

Expanding it up to the first order in the external $p_a$, after the Wick rotation and with use of the Feynman representation we have
\bea
\Sigma_1&=&-\frac{4g^2i}{5}\int_0^1dx\int\frac{d^5l_E}{(2\pi)^5}\frac{l^2_E}{(l^2_E+m^2x)^2}\bar{\psi}(-p)\epsilon^{abcde}\epsilon^{a'b'c'd'e'}\sigma_{de}\gamma^f\sigma_{d'e'}\psi(p)\times\nonumber\\&\times&\eta_{cc'}b_{a}b_{a'}(\eta_{bf}p_{b'}x+\eta_{b'f}p_bx-\eta_{bb'}p_f(1-x)).
\eea
With use of the integrals (\ref{list}), we can rewrite this expression as
\bea
\Sigma_1&=&
-\frac{4g^2i}{5}I_6 \bar{\psi}(-p)\epsilon^{abcde}\epsilon^{a'b'c'd'e'}\sigma_{de}
(\gamma_bp_{b'}x+\gamma_{b'}p_bx-(\gamma\cdot p) \eta_{bb'}(1-x))\sigma_{d'e'}\psi(p)\times\nonumber\\&\times&
\eta_{cc'}b_{a}b_{a'}.
\eea
We substitute the explicit value of $I_6$ and rest with
\bea
\Sigma_1&=&\frac{im^3}{60\pi^2}g^2\int_0^1dx
x^{3/2}\times\nonumber\\ &\times& \bar{\psi}(-p)\epsilon^{abcde}\epsilon^{a'b'c'd'e'}\sigma_{de}
(\gamma_bp_{b'}x+\gamma_{b'}p_bx-(\gamma\cdot p) \eta_{bb'}(1-x))\sigma_{d'e'}\psi(p)
\eta_{cc'}b_{a}b_{a'}+\nonumber\\ &+&\int_0^1dx\left[\frac{4i}{5}\frac{g^2\pi}{2L^3}G_2(\xi\sqrt{x})-\frac{i}{5}\frac{g^2m^2\pi}{2L}xF_2(\xi \sqrt{x})\right]\times\nonumber\\&\times&
\bar{\psi}(-p)\epsilon^{abcde}\epsilon^{a'b'c'd'e'}\sigma_{de}
(\gamma_bp_{b'}x+\gamma_{b'}p_bx-(\gamma\cdot p) \eta_{bb'}(1-x))\sigma_{d'e'}\psi(p)
\eta_{cc'}b_{a}b_{a'}.
\eea
Then we integrate over $x$ in the zero-temperature term:
\bea
\Sigma_1&=&\frac{im^3g^2}{210\pi^2} \bar{\psi}(-p)\epsilon^{abcde}\epsilon^{a'b'c'd'e'}\sigma_{de}
(\gamma_bp_{b'}+\gamma_{b'}p_b-\frac{2}{5}(\gamma\cdot p) \eta_{bb'})\sigma_{d'e'}\psi(p)
\eta_{cc'}b_{a}b_{a'}+\nonumber\\
&+&\frac{ig^2}{5}\int_0^1dx\left[2\frac{\pi}{L^3}G_2(\xi\sqrt{x})-\frac{m^2\pi}{2L}xF_2(\xi\sqrt{x})\right]\times\nonumber\\ &\times&
\bar{\psi}(-p)\epsilon^{abcde}\epsilon^{a'b'c'd'e'}\sigma_{de}
(\gamma_bp_{b'}x+\gamma_{b'}p_bx-(\gamma\cdot p) \eta_{bb'}(1-x))\sigma_{d'e'}\psi(p)
\eta_{cc'}b_{a}b_{a'}.
\eea
The matrix product can be simplified even more.  
Indeed, one can show that
\begin{eqnarray}
&&\eta_{cc'}b_{a}b_{a'}\epsilon^{abcde}\epsilon^{a'b'c'd'e'}\sigma_{de}(a_1(\gamma_bp_{b'}+\gamma_{b'}p_b)+a_2(\gamma\cdot p) \eta_{bb'})\sigma_{d'e'}\nonumber\\ &&=(-72a_1+48a_2)\slashed{b}(b\cdot p)+72a_1\slashed{p}b^2.
\end{eqnarray}
Thus, by considering $b_mb^m=0$ (for light-like case), we can rewrite our result as
\begin{eqnarray}
\label{Si1}
\Sigma_1&=&-\Big[\frac{76im^3g^2}{175\pi^2}+\frac{24ig^2}{5}\frac{\pi}{L^3}\int_0^1dx[2G_2(\xi\sqrt{x})-2\pi^2\xi^2 xF_2(\xi\sqrt{x})](2+x)\Big] \nonumber\\
&&\times \bar{\psi}(-p)\bs(b\cdot p)\psi(p).
\end{eqnarray}
So, we succeeded to generate the aether-like term for the spinor field. We note that at $L\to 0$ the contribution of the compact dimension increases strongly. We note that at the small extra dimension limit $L\to 0$, or, as is the same, $\xi\to 0$, the $F_2$ term will be suppressed in comparison with $G_2$ since in this limit, $2G_2(\xi\sqrt{x})\to-2.4$, and $\xi^2F_2(\xi\sqrt{x})\to0$.

We see that we have both Dirac-like kinetic term (that one proportional to $b^ab_a$) and the aether-like term for the spinor field.

\begin{figure}[htbp]
\centerline{\includegraphics{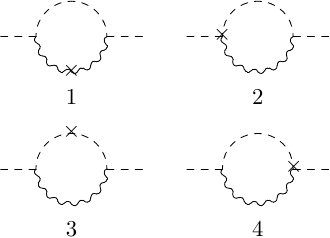}} 
\caption{Contribution to the two-point function of the spinor field.}
\label{F4}
\end{figure}

If we start with the CPT-even theory (\ref{even}), the spinor sector in this case yields the contributions given by Fig.~\ref{F4}.

\vspace*{2mm}

\vspace*{2mm}

As a result, these graphs yield the following first-order contributions to the effective Lagrangian:
\bea
\Sigma_{2,1}&=&4e^2\int\frac{d^5k}{(2\pi)^5}\bar{\psi}(p)\gamma^a
\frac{1}{\ks+\ps-m}\gamma^b\psi(-p)\frac{\kappa_{malb}k^mk^l }{k^4};\nonumber\\
\Sigma_{2,2}&=&-e^2\int\frac{d^5k}{(2\pi)^5}\bar{\psi}(p)\gamma_kc^{ka}
\frac{1}{\ks+\ps-m}\gamma^b\psi(-p)\frac{\eta_{ab}}{k^2};
\nonumber\\
\Sigma_{2,3}&=&e^2\int\frac{d^5k}{(2\pi)^5}\bar{\psi}(p)\gamma^a\frac{1}{\ks+\ps-m}c^{mn}\gamma_m(k_n+p_n)\frac{1}{\ks+\ps-m}\gamma^b\psi(-p)\frac{\eta_{ab}}{k^2};
\nonumber\\
\Sigma_{2,4}&=&-e^2\int\frac{d^5k}{(2\pi)^5}\bar{\psi}(p)\gamma^a
\frac{1}{\ks+\ps-m}\gamma_{\kappa}c^{kb}\psi(-p)\frac{\eta_{ab}}{k^2},
\eea
with $\Sigma_2=\Sigma_{2,1}+\Sigma_{2,2}+\Sigma_{2,3}+\Sigma_{2,4}$. Up to the first order in external momenta, we have
\bea
\Sigma_{2,1}&=&\frac{8}{5}ie^2\kappa_{malb}\int\frac{d^5l_E}{(2\pi)^5}\int_0^1dx(1-x)\frac{l^2_E}{(l^2_E+m^2x)^3}\times\nonumber\\&\times&
\bar{\psi}(p)
\Big[\gamma^a(\ps(1-x)+m)\gamma^b\eta^{ml}-\gamma^a\gamma^m\gamma^bp^l-\gamma^a\gamma^l\gamma^bp^m
\Big]\psi(-p);\nonumber\\
\Sigma_{2,2}+\Sigma_{2,4}&=&-2ie^2\int\frac{d^5l_E}{(2\pi)^5}\int_0^1dx(1-x)\frac{1}{(l^2_E+m^2x)^2}
\times\nonumber\\&\times&\bar{\psi}(p)p^l[-\gamma_lc^a_a+(c_{la}+c_{al})\gamma^a
]\psi(-p)-\nonumber\\
&-&2ie^2mc^a_a\int\frac{d^5l_E}{(2\pi)^5}\int_0^1dx\frac{1}{(l^2_E+m^2x)^2}\bar{\psi}(p)\psi(-p)
;\nonumber\\
\Sigma_{2,3}&=&ie^2\Big\{\int\frac{d^5l_E}{(2\pi)^5}\int_0^1dxx\frac{l^2_E}{(l^2_E+m^2x)^3}\times\nonumber\\&\times&
\bar{\psi}(p)(-\frac{12}{5}c^{mn}(1-x)(\gamma_mp_n+\gamma_np_m)+4mc^m_m-\frac{12}{5}\ps(1-x)c^m_m)\psi(-p)+\nonumber\\&+&
\bar{\psi}(p)\int\frac{d^5l_E}{(2\pi)^5}\int_0^1dx\frac{3}{(l^2_E+m^2x)^2}c^{mn}(1-x)\gamma_mp_n\psi(-p)\Big\}.
\eea
Then, by using the integrals (\ref{list}), we get
\bea
\label{thegamma}
\Sigma_{2,1}&=&\frac{8}{5}ie^2\kappa_{malb}\int_0^1dx(1-x)I_5(m^2x)\times\nonumber\\&\times&
\bar{\psi}(p)
\Big[\gamma^{\alpha}(\ps(1-x)+m)\gamma^b\eta^{ml}-\gamma^a\gamma^m\gamma^bp^l-\gamma^a\gamma^l\gamma^bp^m
\Big]\psi(-p);\nonumber\\
\Sigma_{2,2}+\Sigma_{2,4}&=&-2ie^2\int_0^1dx(1-x)I_1(m^2x)
\bar{\psi}(p)p^l[-\gamma_lc^a_a+(c_{la}+c_{al})\gamma^a
]\psi(-p)-\nonumber\\
&-&2ie^2mc^a_a\int_0^1dxI_1(m^2x)\bar{\psi}(p)\psi(-p)
;\nonumber\\
\Sigma_{2,3}&=&ie^2\Big\{\int_0^1dxxI_5(m^2x)\times\nonumber\\&\times&
\bar{\psi}(p)(-\frac{12}{5}c^{mn}(1-x)(\gamma_mp_n+\gamma_np_m)+4mc^m_m-\frac{12}{5}\ps(1-x)c^m_m)\psi(-p)+\nonumber\\&+&
3\bar{\psi}(p)\int_0^1 dx (1-x)I_1(m^2x)c^{mn}\gamma_mp_n\psi(-p)\Big\}.
\eea
Afterward, the expressions for $I_1$ and $I_5$ must be substituted into the above expression (\ref{thegamma}). Unfortunately, the integral over $x$ from the temperature dependent part are rather complicated. We can simplify these expressions in the special case $c^{mn}=\alpha u^mu^n$, and $\kappa_{mnpq}=\beta(\eta_{mp}u_nu_q-\eta_{mq}u_nu_p+\eta_{nq}u_mu_p-\eta_{np}u_mu_q)$, assuming $u_mu^m=0$ (light-like case). So, we obtain
	\bea
	\label{thegamma1}
	\Sigma_{2,1}&=&-\frac{8}{5}e^2\beta\int_0^1dx(1-x)(-4x+12)\Big[
	-\frac{3m\sqrt{x}}{32\pi^2}+\frac{1}{8\pi L}F_2(\xi\sqrt{x})
	\Big]
	\times\nonumber\\&\times&
	\bar{\psi}(p)
	\us(u\cdot p)\psi(-p);\nonumber\\
	\Sigma_{2,2}+\Sigma_{2,4}&=&4e^2\alpha\int_0^1dx(1-x)
	\Big[
	-\frac{m\sqrt{x}}{16\pi^2}+\frac{1}{8\pi L}F_2(\xi\sqrt{x})
	\Big]
	\times\nonumber\\&\times&
	\bar{\psi}(p)\us(u\cdot p)\psi(-p);\nonumber\\
	\Sigma_{2,3}&=&-e^2\alpha\Big\{\int_0^1dxx(1-x)
	\Big(
	\frac{21m\sqrt{x}}{80\pi^2}+\frac{9}{40\pi L}F_2(\xi\sqrt{x})
	\Big)\Big\}\times\nonumber\\&\times&
	\bar{\psi}(p)
	\us(u\cdot p)\psi(-p).
	\eea
The complete result for the two-point function of the spinor field yields:
\bea
\Sigma_2(p)&=&e^2\bar{\psi}(p)
	\us(u\cdot p)\psi(-p)\times\nonumber\\&\times&
 \Big[-\frac{8}{5}\beta
 \int_0^1dx(1-x)(-4x+12)\big[
	-\frac{3m\sqrt{x}}{32\pi^2}+\frac{1}{8\pi L}F_2(\xi\sqrt{x})
	\big]+\nonumber\\&+&
 4\alpha\int_0^1dx(1-x)
	\big[
	-\frac{m\sqrt{x}}{16\pi^2}+\frac{1}{8\pi L}F_2(\xi\sqrt{x})
	\big]-\nonumber\\&-&
 \alpha\int_0^1dxx(1-x)
	\big[
	\frac{21m\sqrt{x}}{80\pi^2}+\frac{9}{40\pi L}F_2(\xi\sqrt{x})
	\big]
 \Big].
\eea
Again, we find that the aether term arises for the spinor field. Moreover, the $L$-dependent terms dominate as $L$ tends to zero.

\section{Summary}

We calculated the aether-like terms in a five-dimensional gauge theory with magnetic coupling in the case when the extra dimension is compact, generalizing thus the results of \cite{aether}. Actually, we performed summation over the Kaluza-Klein tower of fields, whose methodology was similar to the finite-temperature calculations. We found that, besides the usual aether term and the finite renormalization of the Maxwell term, we have some new terms which can be treated as certain five-dimensional generalizations of the aether-like terms $\kappa^{mnpq}F_{mn}F_{pq}$, which in four dimensions are treated as an important ingredient of the Lorentz-breaking extension of the standard model \cite{Kostel}. These new terms become dominant in the limit where the extra dimension is very small.  This allows us to suggest that there will be essential modifications of quantum corrections if our space indeed has a small extra spatial dimension. Moreover, in the CPT-even case, we have the extra particles described by $A_5$. 

Besides this, we demonstrate the arising of an aether-like term of the standard form \cite{Carr} for the spinor field. Its impact increases as the extra dimension becomes less. 

To close the paper, we emphasize that the presence of the compact extra dimension will essentially modify the observed values of fields and couplings more, as the extra dimension is smaller, which probably could be a crucial fact for the hierarchy problem.

\textbf{Acknowledgments.} The authors are grateful to K. Bakke for important discussions. This work was partially supported by Conselho
Nacional de Desenvolvimento Científico e Tecnol\'{o}gico (CNPq). The work by A. Yu. P. has been supported by the CNPq project No. 301562/2019-9. JF would like to thank the Fundação Cearense de Apoio ao Desenvolvimento Cient\'{i}fico e Tecnol\'{o}gico (FUNCAP) under the grant PRONEM PNE0112-00085.01.00/16 for financial support.


\begin{thebibliography}{50}
\bibitem{KK} Th. Kaluza,  Sitzungsber. Preuss. Akad. Wiss. Berlin. (Math. Phys.), 966–972 (1921).
\bibitem{KK1} O. Klein, Z. Phys. A 37 (12). 895–906 (1926).
\bibitem{BL} D. Bailin, A. Love, Rep. Prog. Phys. 50, 1087 (1987).
\bibitem{ArkHam} N.~Arkani-Hamed, A.~G.~Cohen and H.~Georgi,
Phys. Rev. Lett. \textbf{86}, 4757 (2001)
[arXiv:hep-th/0104005 [hep-th]].
\bibitem{Carr} S. Carroll, H. Tam, Phys. Rev. D78, 044047 (2008), arXiv: 0802.0521.
\bibitem{Rizzo} T. Rizzo, JHEP 0509, 035 (2005), hep-ph/0506056.
\bibitem{Yamashita} N. Maru, T. Yamashita, Nucl. Phys. B754, 127 (2006), hep-ph/0603237.
\bibitem{CST} J.~F.~Assun\c{c}\~ao, T.~Mariz, J.~R.~Nascimento and A.~Y.~Petrov,
JHEP 08, 072 (2018),
arXiv:1805.11049.
\bibitem{Obousy} R. Obousy, G. Cleaver, Mod. Phys. Lett. A24, 1495 (2009), arXiv: 0805.0019.
\bibitem{aether} M. Gomes, J. R. Nascimento, A. Yu. Petrov, A. J. da Silva, Phys. Rev. D81, 045018 (2010), arXiv: 0911.3548.
\bibitem{Maluf} R. V. Maluf, T. Mariz, J. R. Nascimento, A. Yu. Petrov, Int. J. Mod. Phys. A33, 1850018 (2018), arXiv: 1604.06647.
\bibitem{Scarp} A. P. Baeta Scarpelli, T. Mariz, J. R. Nascimento, A. Yu. Petrov, Eur. Phys. J. C73, 2526 (2013), arXiv: 1304.2256.
\bibitem{Ford} L. H. Ford, Phys. Rev. D21, 933 (1980).
\bibitem{KostMew} V. A. Kostelecky, M. Mewes, Phys.\ Rev.\ D88, no. 9, 096006 (2013),
arXiv:1308.4973 [hep-ph].
\bibitem{Kostel} V. A. Kostelecky, Phys. Rev. D69, 105009 (2004), hep-th/0312310.

\end{thebibliography}
\end{document}